# LARGE-SCALE CORBA-DISTRIBUTED SOFTWARE FRAMEWORK FOR NIF CONTROLS

Robert W. Carey, Kirby W. Fong, Randy J. Sanchez, Joseph D. Tappero, John P. Woodruff
Lawrence Livermore National Laboratory, Livermore, CA 94550, USA

Abstract

The Integrated Computer Control System (ICCS) is based on a scalable software framework that is distributed over some 325 computers throughout the NIF facility. The framework provides templates and services at multiple levels of abstraction for the construction of software applications that communicate via CORBA (Common Object Request Broker Architecture). Various forms of object-oriented software design patterns are implemented as templates to be extended by application software. Developers extend the framework base classes to model the numerous physical control points, thereby sharing the functionality defined by the base classes. About 56,000 software objects each individually addressed through CORBA are to be created in the complete ICCS. Most objects have a persistent state that is initialized at system start-up and stored in a database. Additional framework services are provided by centralized server programs that implement events, alerts, reservations, message logging, database/file persistence, name services, and process management. The ICCS software framework approach allows for efficient construction of a software system that supports a large number of distributed control points representing a complex control application.

## 1 INTRODUCTION

The ICCS is a distributed, hierarchically organized, object-oriented control system that employs a framework of reusable software to build uniform programs to satisfy numerous functional requirements. ICCS employs Ada, CORBA, and object-oriented techniques to enhance the openness of the architecture and portability of the software. Java is used for the production of graphical user interfaces and the integration of commercial software, particularly the Oracle database system.

NIF is a largely repeating structure of 192 laser beams, each having almost identical sets of components. The application software elements are representations of that laser hardware devised to map NIF physical control components to CORBA software objects [1][2][3].

Following strategies of object-oriented software development, similar software components are defined as classes, and these classes are instantiated for each occurrence of a NIF component in each laser beam. The framework and application software combined have resulted in a design consisting of approximately 300 interface classes.

Each software component derives from a particular component class and has an identity (name) within the facility. Each software component object is allocated to the processor that is connected to its physical representation. CORBA maps the name to the software component object located in the processor that is connected to the physical hardware. This gives location transparency to the ICCS application software.

## 2 LAYERED ARCHITECTURE

The ICCS layered architecture [4] was devised to address the general problem of providing distributed control for large scientific facilities that do not require real-time capability within the supervisory software. The resultant architecture consists of front-end processors (FEPs) coordinated by a supervisory system and exploits commercially available components where possible.

Functions that operate hardware control points are implemented on FEPs. There are 18 different types of FEP computers – some 300 computers in all – that differ by the hardware devices that they control. The control points themselves are sensors and actuators attached to interface boards plugged into an FEP backplane. In many cases, control points are handled by intelligent components that incorporate embedded controllers operated by small fixed programs. This firmware running in the embedded controller does much of the low-level work that would otherwise be allocated to an FEP.

The supervisor layer is partitioned into ten cohesive subsystems [5], each of which controls a primary NIF subsystem such as beam control or power conditioning. These supervisory controls, which are hosted on UNIX workstations housed in control consoles, provide centralized operator controls and status, data archiving, and integration services.

Some real-time control is inevitably necessary. Functions requiring real-time implementation are allocated to software within a single FEP or to an embedded controller, so communication over the local area network is not obligated to meet hard-deadline schedules.

## 3 SOFTWARE FRAMEWORKS

The ICCS software framework is the collection of collaborating abstractions that are used to construct the application software. This framework delivers prebuilt components that are extended to accommodate specific additional requirements in the construction of the application software. Engineers specialize the framework for each subsystem [6] to handle different kinds of control points, controllers, user interfaces, and functionality. The framework concept enables the cost-effective construction of the NIF software and provides the basis for long-term maintainability and upgrades. The following discussion introduces the framework components that form the basis of the ICCS software.

**System Manager** – provides services for the integrated management of the ICCS network of hundreds of computers, ensures that the necessary processes and computers are operating, and provides failover services for processes that terminate during operation.

**Graphical User Interface (GUI)** – enables human interaction with the ICCS via graphical user interfaces displayed upon control room consoles or on X Terminals distributed throughout the facility. The GUI is implemented as a framework to ensure consistency across the applications. Commercial Java-based GUI development tools are used to construct the display graphics.

**Message Log** – provides audit trace services for all subsystems. A central server collects messages and associated attributes and writes them to appropriate persistent stores. Interested observers use Database Management System retrieval techniques to update operators' consoles or to produce historic audit trails of operations.

**Configuration** – defines the naming convention and manages the static data for the hardware control points that are accessible to the ICCS. Configuration provides a taxonomic system that is used as the key by which clients locate software control components on the CORBA bus. Coupled with the use of the Factory pattern, the framework application allows processes to postpone until runtime the particular arrangement of equipment that is allocated to individual computers on the network.

**Application** – defines templates for aggregation, iteration, and control of collections of control components. A publish/subscribe mechanism that implements the Observer pattern provides a means for the GUI layer to interact with the control layers. This framework is the basis for messages that transmit status updates between the Ada and Java languages using data-type independent containers.

**Reservation** – manages access to control points by giving a single client exclusive rights to control or otherwise alter the control point. The framework uses a lock-and-key model. Reserved control points that are "locked" can only be manipulated if and when a client presents the "key."

**Status Monitor** – provides generalized services for disseminating control point status information using the push model of change notification. The status monitor operates within the FEP, observing control points and notifying other parts of the system when the status changes by a significant amount. Network messages are only generated when changes of interest occur.

**Alert System** – any application encountering a situation that requires immediate attention raises an alert, which then requires interaction with an operator to proceed. The alert system records its transactions so that the data can be analyzed after the fact.

**Machine History** – gathers information about equipment performance during operation of the NIF for analysis to improve efficiency and reliability. Examples of such information are component installation and service records, operating service time or usage count, abnormal conditions, and periodic readings of sensors.

**Sequence Control Language** – used to create custom scripting languages for the NIF applications. The service automates sequences of commands executed on the distributed control points or other software artifacts. The implementation is based on Extended Markup Language (XML) data language technology and has been used to provide operator-selectable image processing algorithms in an automated beam alignment application.

**Shot Life Cycle** – defines a finite state machine relating the nine intermediate states in preparing for, firing, and analyzing a NIF shot. The framework provides coordination, and the several functional subsystems define the details of work to be done at each of the stages of shot execution.

**Shot Archive** – provides the repository services for data retrieved from the various instruments and sensors that diagnose performance of the laser and target interactions.

## 4 LEVELS OF ABSTRACTION

In a large system, coupling between software components can be managed by dividing the system into subsystems that are organized by levels of abstraction, and then enforcing rules for dependencies between subsystems.

The discipline of layers of abstraction prescribes that every component within an upper-level subsystem may depend only on components that are allocated to a specified set of lower-level subsystems. The upper subsystem is said to "import" the lower subsystem. Cycles in the graph of imports are forbidden, but within a subsystem, packages may mutually depend on each other.

There are two benefits to organizing components into leveled subsystems. The effect of changes to an interface is limited to the components contained in subsystems that import the changed component. In addition, developers working on low-level packages can enhance and test their product and release stable subsystems for upper-level work, according to a managed schedule without inconveniencing application developers who import prior versions [7].

In object-oriented programming, abstractions are extended by deriving new types, and those derivations may cross subsystem boundaries. Levels of abstraction have caused ICCS packages to be allocated into subsystems at three levels called Framework Templates, NIF Building Blocks, and Application Behaviors. The subsystems in these three levels provide a low level of reusable templates, an intermediate level of concrete components, and an upper level that instantiates the templates into services that act on the components. All the subsystems adhere strictly to the principle limiting imports to lower levels.

Subsystems in the Framework Templates level provide abstractions that arise from a domain analysis of control systems in experimental facilities. Control systems operate on control points denoted as "devices" in the ICCS. ICCS defines a class rooted at an abstract type to represent devices. This base type declares properties to be shared by all the control points implemented in the FEPs: they possess references allowing distributed access via CORBA, they are initialized with data from a central data store, and they can be reserved to assure exclusive operation by a single client. Facility for monitoring and publishing their status is in this level as well. However, this facility is incomplete (in the templates level) since the service is defined in terms of the abstract type.

The NIF Building Block level contains an inheritance hierarchy that extends the abstraction of the device to implement all the diverse kinds of procedures that real physical devices – motors, power supplies, transient digitizers, and the like – provide for their users. The tactics of inheritance and aggregation are both used to define objects in this level. These extensions reify the interfaces that were needed to fulfill the domain analysis. Therefore a motor device can report its present position to the monitoring framework, and the framework publishes that status.

A complete set of control system functions is built on the levels above the Building Blocks so the services remain available when the device class is extended. Application Behavior, the uppermost of the three layers, aggregates building blocks and extends the services promised by the Templates layer. The concrete packages that are defined here are extensible because they use the polymorphic types in the building block layer. And these packages, once extended, are the components from which the ICCS Main Programs in the highest-level subsystems are built.

## 5 PATTERNS

Use of existing object-oriented patterns is a proven practice for constructing robust software systems [8]. Below is a brief discussion of some of the more prevalent design patterns in the ICCS distributed architecture.

**Observer** – The ICCS Framework makes extensive use of the observer pattern (publish/subscribe) which is well suited to a distributed environment because it decouples publishers from knowledge of subscribing clients. The Event, Alert, Status Monitor, and Shot Life Cycle Frameworks all make use of this pattern. In most all cases, the publishers and subscribers are in separate processes and communicate via CORBA.

**Factory** – Object factories are used for the creation and initialization of all application objects in the ICCS. Each process has object factories that are registered with the configuration server at process start-up time and are remotely instructed to create objects that define the content of an application program. Configuration data for each process is stored in a database and served to the object factories via CORBA.

**Model-View-Controller** – The model-view-controller (MVC) architecture is a variation of the observer pattern that serves to decouple the operator interface (GUI layer) from the application software, dividing the appearance of the interface from the control object that defines system semantics. CORBA provides interface technology that allows transparent implementation of the MVC architecture across language environments.

**Strategy** – The purpose of the NIF is execution of laser experiments called "shots." The ICCS utilizes the strategy pattern to allow the "shot" logic for different subsystems comprising the NIF to vary independently of the "shot" state.

**Additional Patterns** – Many structural patterns are used to facilitate various kinds of composition (*adapter, bridge composite, proxy*). Inheritance and support of polymorphic behaviors are the basis for the

ICCS Framework Templates. In addition, other behavioral patterns such as *command, mediator, state*, and *template method* are incorporated in various parts of ICCS framework and application software.

Extending these patterns to a distributed environment adds complexity and requires incorporation of connection management. The ICCS Framework is built on existing design patterns that have been extended for distribution. Distribution presents additional failure modes that must be addressed in the software designs. The implementation of a design pattern must include logic for failure and recovery when portions of the pattern reside on different computers.

## 6 DISTRIBUTED COMPONENT ARCHITECTURE

The NIF physical organization lends itself to a distributed component-based communication architecture. Control components consist of various actuators, sensors, and instruments used to activate and diagnose each NIF laser beam and its interaction with a target. The ICCS maps NIF physical control components to CORBA software objects. The interface to each software component object is defined by the allowable operations that are supported by the physical representation.

This model is an elegant and easy to understand representation of the physical NIF. However, with distribution come challenges for building robust, resilient systems. No software application can be complete until error conditions are handled. A distributed system running on top of the TCP/IP transport layer has many more error possibilities than a monolithic, nondistributed system.

The number of distributed interfaces correlates directly to system complexity. A high number of distributed interfaces allow possibly nondeterministic messaging behavior. The potential for deadlock in such an open communication environment is ever present. Stringent testing can expose many deadlocks, but some will likely arise in production operation. Debugging distributed systems is difficult at best. In addition to deadlock potential, highly distributed systems must deal with connection management. Individual systems might need to be restarted for any of several reasons. How clients deal with lost connections has a direct impact on system resistance to failure.

ICCS has developed standards for interface design that specify various communication decoupling mechanisms[6]. These decoupling mechanisms are integral to the designs of the distributed software framework and provide examples/patterns that application interfaces can employ. The ICCS Framework also contains connection abstractions that manage the health of a CORBA reference.

## 7 PERSISTENCE LAYER

The ICCS architecture employs server programs that act as persistence brokers [9] to provide database or file services to thousands of distributed objects on hundreds of computers. The brokers provide the run-time interface between the control system and the persistence mechanism while hiding the persistence mechanism from the application software. CORBA allows variation of the implementation language and persistence mechanism independently of the application software, as was the case with GUI technology. ICCS currently has persistence brokers that interface to a relational database (Oracle), to XML, and to HDF (Hierarchical Data Format) files.

ICCS persistence classes map the data portions of object designs to a specific persistence format. Addition of new classes and their associated persistence is easily accomplished. Base class templates are provided to build the specific persistent subclasses. These base classes are extended through inheritance by application programmers to implement specific persistent behaviors.

Each persistence broker can support one or more persistent base classes. The broker programs are constructed as semi-stateless processes. The state information that the persistence brokers require is maintained in the database. As the state information changes, the database is updated. This allows us to start/restart programs as necessary without having to relearn the state of the control system.

## 8 PROCESS MANAGEMENT

Inter-process dependencies among a large population of processes introduce substantial complexity. The ICCS system manager framework maintains rudimentary process state information from heartbeats. Since it will be a typical occurrence for some computers to be stopped and restarted for a variety of reasons, reliable connection management requires that clients be notified when the state of a service in another process changes. The challenge is to provide client notification of server state changes while maintaining the location transparency in the application software.

Plans for enhancing the process management mechanism in ICCS will enable observation of resource loading and monitoring and control of process states. Statistics on the state of network components, CPU utilization, memory consumption, file system capacity, and message traffic are important for managing the health of ICCS.

Existing network management technology is being leveraged to support ICCS process management. HP-OpenView and Simple Network Management Protocol (SNMP) were chosen as the process management tools of choice, and investigation of existing SNMP agents for the target platforms has been started. Some customization of agents and MIBs (Management Information Bases) will be required to customize HP-OpenView and SNMP to the ICCS environment. Research into the capability to have SNMP agents monitor the object request broker is beginning. There is significant work remaining in this area of the ICCS architecture.

## 9 SUMMARY

In the context of a distributed system design, choices for the ICCS component-based communication architecture map to the physical organization of the NIF. A layered architecture has been employed to manage dependencies between different levels of abstraction. The ICCS supervisory software framework delivers prebuilt components that are extended to accommodate specific additional requirements in the construction of the application software. Twelve different framework abstractions are described that provide the basis for constructing the NIF Control System. Object-oriented design patterns are extended to support reliable distribution. The ICCS persistence layer has been successful encapsulating a variety of persistence mechanisms.

Construction of the application software is progressing in a planned, iterative fashion using the software framework foundation layers. Process management features are being designed and added to the framework in preparation for scaling to the full ICCS process population. Distribution complexity and connection management continue to be the challenge to the software architecture as the number of control points and FEPs are scaled to the full NIF.